\documentclass[journal]{IEEEtran}

\usepackage{graphicx}
\usepackage{subcaption}
\newtheorem{exa}{Example}

\ifCLASSINFOpdf

\else

\fi

\begin{document}

\title{Uniform Array with Broadband Beamforming for Arbitrary Beam Patterns}

\author{Phan~Le~Son
	\thanks{The author is with the Department of Electrical \& Computer Engineering, Technische Universität Kaiserslautern, Kaiserslautern, Germany. Email:phan@eit.uni-kl.de}}

\maketitle

\begin{abstract}
Broadband beamforming is a technique to obtain the signal with a wide range of frequencies. It maintains the signal integrity and spatial selectivity over
frequencies. This is important in several applications such as microphone array, sonar array or radar where the operation range of the signal is several octaves.
Using uniform array for broadband beamforming is the old topic but there is no available design method for the arbitrary beam patterns, except the optimization methods. In this paper, we present a new method based on geometry translation and coordinate transformation to design broadband beamformer for arbitrary beam patterns. The new method uses less computation time than the optimization methods and it could help to find a better configuration of the array such as fewer sensors, smaller size.
\end{abstract}


\IEEEpeerreviewmaketitle

\section{Introduction}
 Beamforming is a common technique used in array of sensor for directional signal reception. It applies the spatial filter to obtain the beam pattern. The array with fixed inter distance of the sensor works well with the narrowband signal \cite{CDB:Darren}. However, in real applications, many systems need a wide band of frequency, such as the audio signal has a wide band of frequency, from 20 Hz to 20 kHz for high-quality recording. There is a high demand for broadband beamformer.

There have been several techniques proposed to overcome this problem, such as Optimization \cite{Broadband:Sven}, harmonic nesting \cite{Autodirective:JL,MA:McCowan,MA:Kellermann,MA:Khalil}, exploiting the Fourier transform relationship between the array’s spatial and temporal parameters and its beam pattern \cite{Broadband:Wei,broadband new:Liu,CDB:Darren}, etc. The studies mention in \cite{Broadband:Wei,CDB:Darren} are effective methods, they utilize the relationship between the beam pattern and Fourier transform but they still have some limitations. The method proposed by Wei Liu \cite{Broadband:Wei} is not so flexible since the directivity function is a function of $cos, sin$ of direction angles. It does not mention the relationship between the working frequency ranges and array configurations. The method proposed by Darren B.Ward \cite{CDB:Darren} is difficult to use, it restricts the weight's complex value of sensor is proportional to the frequency. The optimization methods \cite{Broadband:Sven} work for arbitrary beam pattern but it needs to randomly select the parameter for optimization which is time-consuming and it is difficult to find the optimal configuration for the array. Sometimes, we may need more sensors than they require. 

In this study, we propose a new way to obtain the weight's complex value for arbitrary beam patterns. We still use the relationship between the beam pattern and the Fourier transform. Additionally, we apply the geometry transforming for the beam pattern. The expected beam pattern is translated to Spherical coordinate and then to Cartesian coordinate before applying Invert Fourier transform. The method is simple but works for any expected beam pattern (even in case the beam pattern is a map of direction angles). Besides, the study suggests the method to evaluate the beam pattern quality depends on frequency, number of sensors, inter-distance of sensor. Hence, the designers are more flexible to adjust the sensor array, it is possible to make the trade-off between extending the working frequency ranges and the beam pattern deformations over frequencies.

The paper is organized as follows. In Section II, we addresse the problem as a continuous sensor. In Section III, we talk about the discrete sensor and three examples are given. Finally, conclusions are drawn in Section IV.

\section{Continuous sensor}

Sensor at position $(x,y)$ in Cartesian coordinate has the value $p(x,y,f)$ at frequency f in frequency domain. And a weight: $w(x,y,f)$ is the filter element.
The output of the sensor array is:
\[{ y(f) \ = \ \int\int_{-\infty}^{\infty} w(x,y,f)p(x,y,f)dxdy}\]
Considering the far-field wave, then the wave is planar. $c$ is the speed of wave propagation. The spatial response for a source at azimuth and elevator angle $(\phi,\theta)$:
\begin{equation}
\label{Int:2D}
b(\theta,\phi,f) = $$ 
 $$\int\int_{-\infty}^{\infty} w(x,y,f)\cdot e^{-j{2\pi f\over c}(x\sin\theta\cos\phi+y\sin\theta\sin\phi)}dxdy  	
\end{equation}
Substitutions:
\begin{equation}
\label{Rad:cons}
R = \frac{f}{c}
\end{equation}
\begin{equation}
\label{Coor:Trans}
{u = R\sin\theta\cos\phi}, \quad {v=R\sin\theta\sin\phi} 
\end{equation}

$b(\theta,\phi,f)$ could be presented in $b(u,v,f)$, drives from (\ref{Int:2D}):
\begin{equation}
\label{BeamUV:Cons}
b(u,v,f) = \int\int_{-\infty}^{\infty} w(x,y,f)\cdot e^{-j{2\pi }(ux+vy)}dxdy
\end{equation}

From equation above, we see $b(u,v,f)$  is 2D Fourier transform of $w(x,y,f)$ with respect to $x,y$ or $w(x,y,f)$  is 2D Inverse Fourier transform of $b(u,v,f)$.

At a single frequency, $R=f/c$ is constant, (\ref{Coor:Trans}) is the formula transforming a surface $b_f(\theta,\phi)$ in Spherical coordinate to a surface $b_f(u,v)$ in Cartesian coordinate. Our method is derived from this statement. We translate the expected beam pattern to the functions in Spherical coordinate and then transforming the function in Spherical coordinate to Cartesian coordinate before applying invert Fourier transform.

\begin{figure}[h!]
	\includegraphics[width=\linewidth]{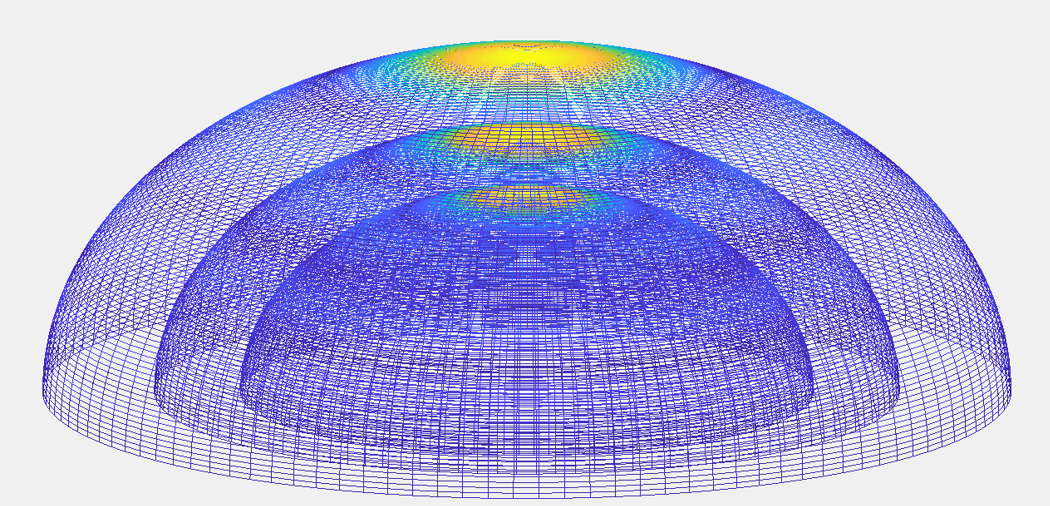}
	\caption{Example of beam-patter presented in Spherical coordinate at 3 different frequencies}
	\label{fig:fig1}
\end{figure}

In Spherical coordinate, radius $R$ is proportional to frequency $f$: for a single value $R$, it is mapping with single value $f$ or vice versa. The expected beam pattern is presented by the function $B(\theta,\phi)$:
\begin{itemize}
\item A beam pattern $B(\theta,\phi)$ is translated to the "gain" functions in Spherical coordinate $b_R(\theta,\phi)$: different radius $R$ associated with different "gain" function presents for different frequency $f$.
\item A "gain" function in Spherical coordinate $b_R (\theta,\phi)$ is located in the haft sphere with radius $R$ (the gain value could be presented by color on the surface of the sphere, Figure.\ref{fig:fig1})
\item $b_R (\theta,\phi)$ is transformed to the function $b_f (u,v)$ in the Cartesian coordinate by formula (\ref{Coor:Trans}). $b_f (u,v)$ is the "gain" function at frequency $f$.
\end{itemize}
(For Line sensor array, the expected beam pattern is able to transform to Polar coordinate and then it is transformed to Cartesian coordinate.)
\\
Below, we could suggest a method to design frequency invariance beam pattern:
\begin{itemize}
\item Step 1: Define the expected beam pattern \\

\item Step 2: For a single R, presenting expected beam pattern to a “gain” function $b_R (\theta,\phi)$ in Spherical coordinate.\\
 
\item Step 3: For a single frequency $f=Rc$, $b_f(u,v)$  is achieved by converting the surface $b_R(\theta,\phi)$ in Spherical coordinator to a plan $(u,v)$ in Cartesian coordinate.\\
 
\item Step 4: Taking the 2D inverse Fourier transform of $b_f(u,v)$ to achieve $w(x,y,f)$. 

\end{itemize}
Note that before and after doing inverse Fourier transform, shifting zero-frequency component to begin and center of spectrum are required.
The weights derive from this solution could form the beam pattern close with the expected beam pattern. The distortion results from the distortion of geometry when translating and transforming, therefore it depends on number of sensors, equidistance of grid sensor and frequency.
\section{Discrete sensor and examples}

\begin{figure}[h!]
	\includegraphics[width=\linewidth]{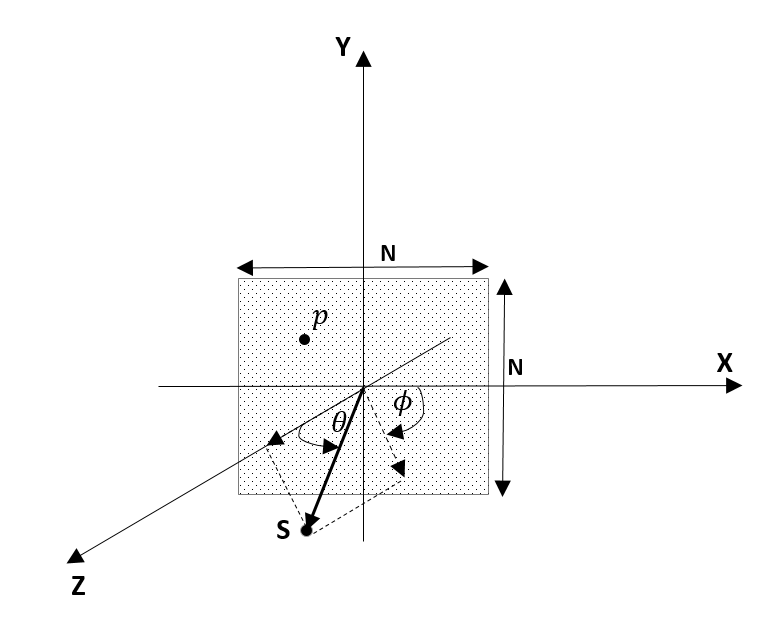}
	\caption{Discrete sensor array in Cartesian coordinate}
	\label{fig:fig0}
\end{figure}

Considering the planar square array: number of sensors in $X$ direction is equal number of sensors in $Y$ direction. 
For the sake of simplicity, we only consider the case: equidistance of inner sensor in $X$ direction is equal with the equidistance in $Y$ direction, the number of sensors in each direction is an odd number. The Origin of coordinate is the center of the planar array (Figure.\ref{fig:fig0}). 
\\
$N$: Number of sensors in $X$ or $Y$ direction (odd number).\\
$d_H$: Equidistance of sensors in $X$ or $Y$ direction.\\
$w(n_1,n_2,f)$: Weight factor of sensor at position $(n_1,n_2)$ ,${-(N-1)\over 2} \leq n_1,n_2\leq {(N-1)\over 2}, n_1,n_2 \in N$. (\ref{Int:2D}) becomes:
\begin{equation}
b(\theta,\phi,f) =  \sum_{n_1,n_2=-{(N-1)\over 2}}^{{(N-1)\over 2}}w(n_1,n_2,f)$$
$$\times e^{-j{2\pi f\over c}(n_1 d_H\sin\theta\cos\phi+n_2 d_H\sin\theta\sin\phi)}
\end{equation}             
We define
\begin{equation}
\label{Coor:DisTrans}
u = {{fNd_H \over c} \sin\theta\cos\phi}, \quad {v \ = {fNd_H \over c}\sin\theta\sin\phi} 
\end{equation}
Radius in Spherical Coordinate is computed as follows:
\begin{equation}
\label{Radius:Dis}
R = {fNd_H \over c} 
\end{equation}                                       
(\ref{BeamUV:Cons}) becomes: \\ 
$
\label{Beam:Dis}
b(u,v,f) \ =  \sum_{n_1,n_2=-{(N-1)\over 2}}^{{(N-1)\over 2}}w(n_1,n_2,f) 
e^{-j{2\pi \over N}(n_1 u +n_2 v)}
$\\
 
 The above equation is a 2D discrete Fourier transform with respect to the variables $n_1,n_2$. We can imagine the variables $n_1,n_2$ build a planar grid points associated with grid points $u,v$ in Cartesian coordinate. Therefore, every grid point in Cartesian coordinate is associated with a sensor. For  discrete sensor, we adopt the design method for continuous sensor by replacing (\ref{Rad:cons}), (\ref{Coor:Trans}) by (\ref{Radius:Dis}), (\ref{Coor:DisTrans}). Invert Fourier transfrom is replaced by Invert Discrete Fourier Transfrom (IDFT).
 
 In general, the range of working frequencies depends on configuration of the grid sensors, the Radius in Spherical coordinate should not be greater than the boundary of the planar sensor array (associated with the grid points) in Cartesian coordinate and at least more than 5 grid points inside the sphere. This constraint ensures the coordinate transformation from Spherical coordinate to Cartesian coordinate is not deformed (loss information).
 $$1 \leq R \leq {(N-1)\over 2}.$$
 Using (\ref{Radius:Dis}), 
 $$1 \leq {fNd_H \over c} \leq {(N-1)\over 2}.$$
 We obtain 
 $${c \over Nd_H} \leq f \leq {c(N-1) \over 2Nd_H}.$$ 
 It is equivalant to the wavelength {$\lambda$ constraint,
 	\[ {Nd_H \over 2} \geq {\lambda \over 2} \geq {Nd_H \over N-1}.\]
 The same approach for N even, we could have the following constraints for frequency:
 \begin{equation}
 \label{Dis:Condition}
 \cases{ {c \over Nd_H} \leq f \leq {c(N-2) \over 2Nd_H}, \;  {\hbox{ N is even}} \  \cr
 	\  \cr
 {c \over Nd_H} \leq f \leq {c(N-1) \over 2Nd_H}, \;  {\hbox{ N is odd}}} 
 \end{equation}
The origin of Catersian coordinate is always consolidated with a sensor's position. In case $N$ is even, it is not at the center of array. Therefore, the maximum of radius is ${(N-2)\over 2}$. However, we could select the origin of Catersian coordinate is at center of array (not associated with any sensor) and we need to do the interpolation when transforming beam pattern from Spherical coordinate to Cartesian coordinate in order to get the values in Cartesian coordinate associated with sensors'postions. In such a case, the maximum of radius is ${(N-1)\over 2}$. For the sake of simplicity, we do not consider the interpolation in this paper.\\    
In practical application, we could extend the working frequency ranges if small deformation is acceptable. This property is applied in "example \ref{Exa:Cylinder}".
For the discrete sensor, we could use (\ref{Coor:DisTrans}), (\ref{Radius:Dis}) for coordinate convert and use (\ref{Dis:Condition}) for evaluating the frequency ranges.\\

\begin{exa}
	\label{Exa:Cylinder}
	Design beam pattern with open-angle of elevator is $\theta \geq \theta_C$ for discrete planar sensor array:
\end{exa}
\begin{itemize}
\item Step 1: Define the expected beam-pattern, Figure.\ref{fig:fig2}.\\ 
Gain 1: for $\theta \geq \theta_C$.\\  
Gain 0: for $\theta < \theta_C$.  \\

	\begin{figure}[h!]
		\includegraphics[width=\linewidth]{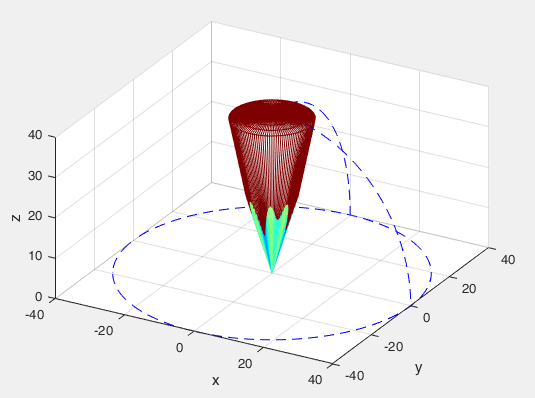}
		\caption{Expected Beam Pattern with open angle $\theta_C$}
		\label{fig:fig2}
	\end{figure}
\item Step 2: 
 For a single R, presenting expected beam pattern to a "gain" function $b_R(\theta,\phi)$ in Spherical coordinator, Figure.\ref{fig:fig3}. \\
$R=const$, we define a surface in spherical coordinate:
\[
b_R( \theta , \phi) \ = {
\cases{1, \; \hbox{$\theta \leq \theta_C$,} \  \cr
0, \; \hbox{otherwise.}
}}
\]
$\theta_C  =\pi/12$: Threshold elevation angle, define the beam-width size. 
	\begin{figure}[h!]
		\includegraphics[width=\linewidth]{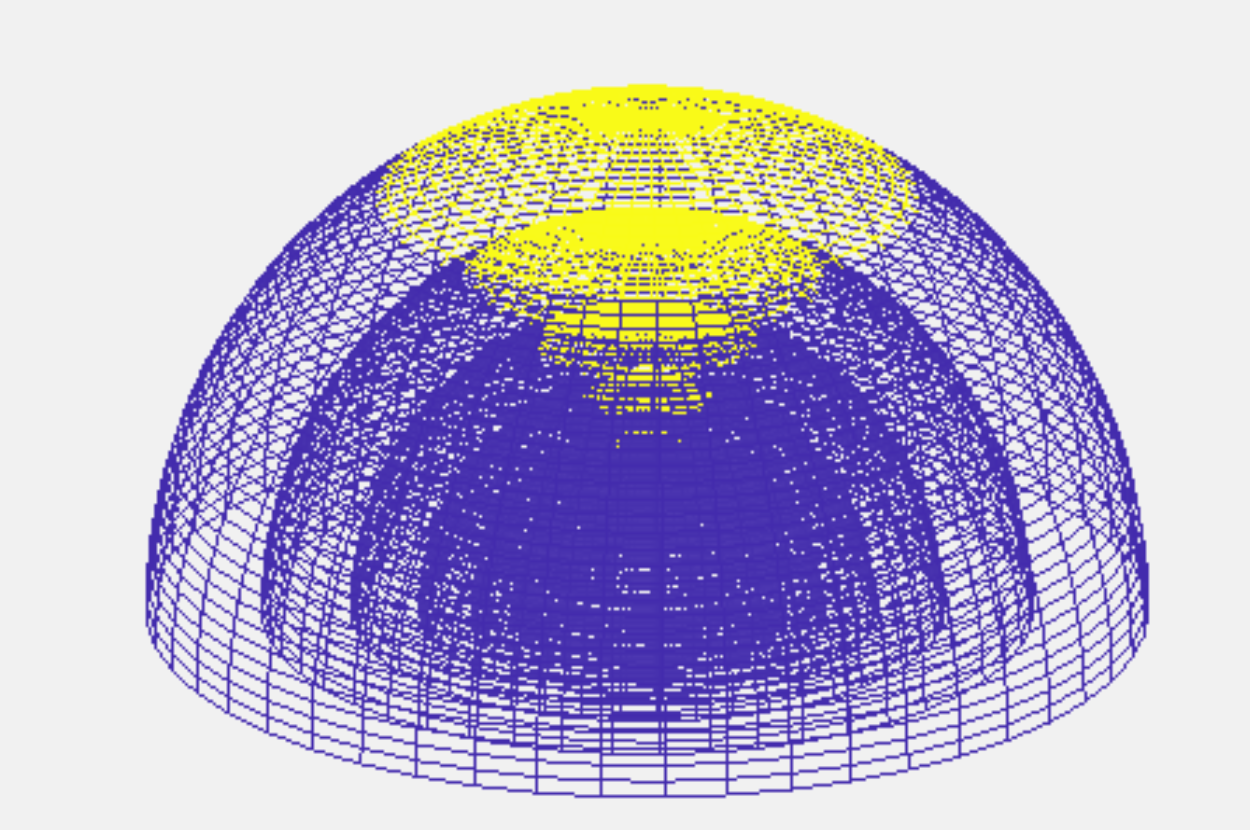}
		\caption{Beam Patterns present in Spherical Coordinate at different Radiuses (frequencies)}
		\label{fig:fig3}
	\end{figure}
\item 	Step 3: 
Convert $b_R(\theta,\phi)$  to Cartesian coordinate ($Rsin\theta= \sqrt{u^2+v^2}$):
\[
b_f( u,v ) \ = {
	\cases{1, \; \hbox{    ${\sqrt{u^2+v^2} \leq {Nd_Hf \over c}\sin\theta_C}$} \  \cr
		0, \;  \hbox{otherwise}
}}
\]
\begin{itemize}
	\item Figure \ref{fig:fig4} is the cylinder shape of "gain" in Cartesian Coordinator. 
	\begin{figure}[h!]
		\includegraphics[width=\linewidth]{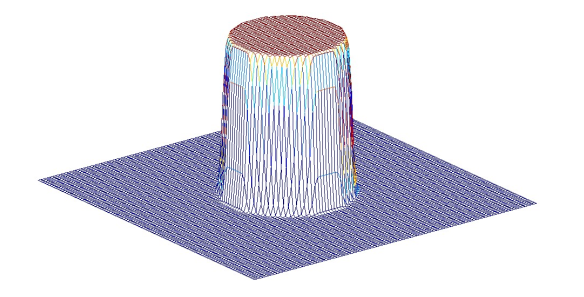}
		\caption{Example of Cylinder at 16Khz, $N=100$, $d_H=0.01 m$}
		\label{fig:fig4}
	\end{figure}
   \item When increasing the frequency, the diameter of the cylinder in Cartesian is expanded.
   \item When decreasing the frequency, the diameter of the cylinder in Cartesian is reduced.
   \item To extend the frequency ranges for this example, we don’t check with the constraints in (\ref{Dis:Condition}). We check the cylinder’s diameter with the boundary.
   \begin{itemize}
   	
   \item If ${Nd_Hf \over c}\sin\theta_C > {N -1 \over 2}$ which means that, 
   $$f > \frac{c(N-1)}{2Nd_H\sin\theta_C}$$ 
   then the cylinder is bigger than the boundary of the array, it leads to the cylinder is distortion. So the constant beam-pattern constraint is not correct anymore.
   \item If ${Nd_Hf \over c}\sin\theta_C < 1$, this implies 
   $$f < \frac{c}{Nd_H\sin\theta_C}$$
   then there is only 1 point in the grid that satisfies the condition to build the cylinder.  
	\end{itemize} 
\end{itemize} 
With $N$=25, $d_H$=0.015m , $\theta_C=\pi/12$, range of frequencies: $3.5 Khz < f < 42.4 Khz$, Figure.\ref{fig:fig5} is the cylinder shape after filtering to avoid the overshoot effect of IDFT.
	\begin{figure}[h!]
		\includegraphics[width=\linewidth]{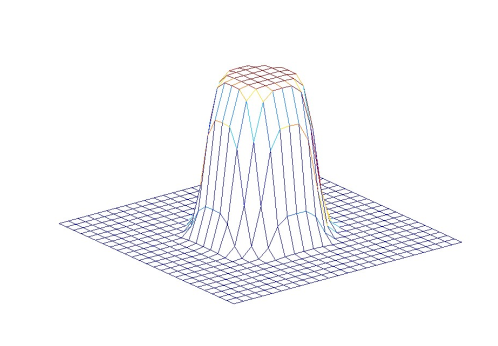}
		\caption{Cylinder shape in Cartesian Coordinator}
		\label{fig:fig5}
	\end{figure}\\
\item Step 4: Taking the 2D IDFT of $b_f (u,v)$ to achieve $w(n_1,n_2,f)$, Figure.\ref{fig:fig6}.\\
	\begin{figure}[h!]
	\includegraphics[width=\linewidth]{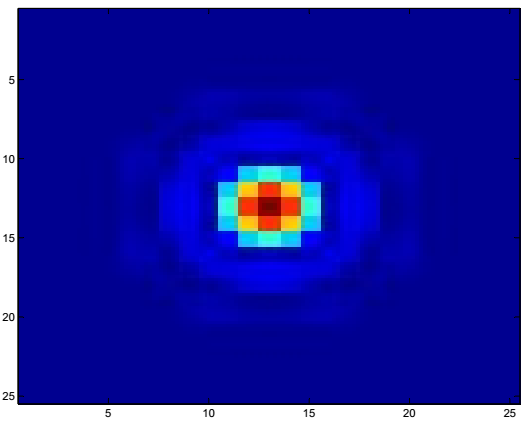}
	\caption{The weight spectrum of 25x25 sensors at 16Khz}
	\label{fig:fig6}
	\end{figure}\\
Beam pattern at 8Khz, 16Khz are showed in Figure.\ref{fig:fig8}.  

	\begin{figure}[h!]
    \centering
    \begin{subfigure}[b]{0.4\linewidth}   		
	\includegraphics[width=\linewidth]{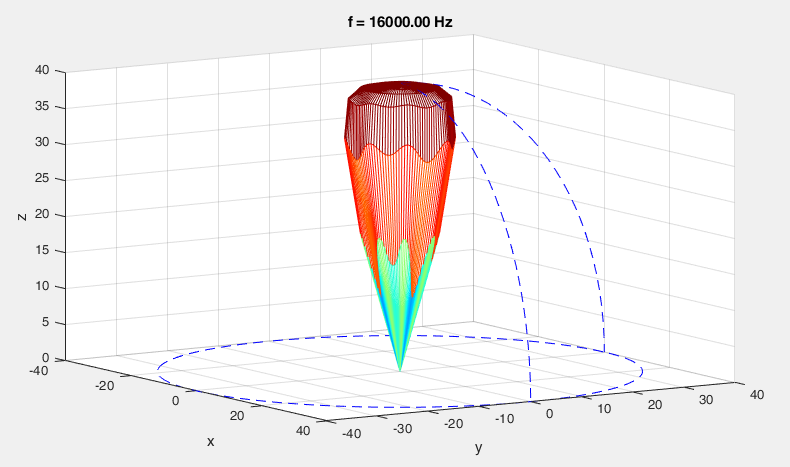}
	\caption{16Khz}
    \end{subfigure}

	\begin{subfigure}[b]{0.4\linewidth}
	\includegraphics[width=\linewidth]{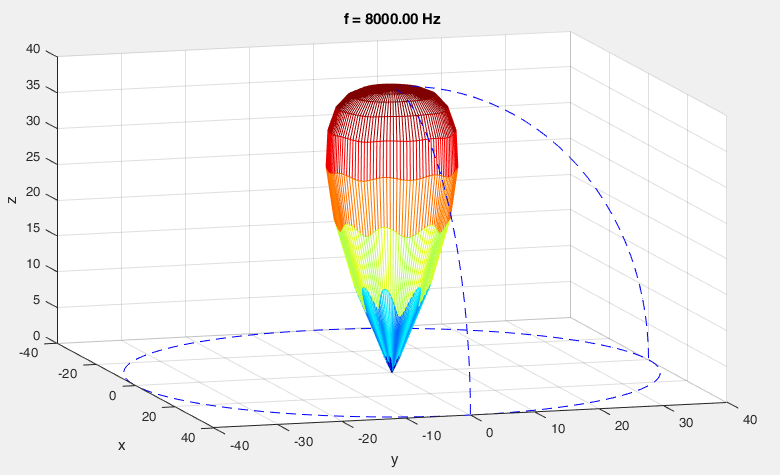}
	\caption{8Khz}
	\end{subfigure}
    \caption{Directivities: The cone shape }
	\label{fig:fig8}
	\end{figure}
Figure.\ref{fig:fig9} presents the cross-cut of beam patern from 1Khz to 16Khz. 
	\begin{figure}[h!]
	\includegraphics[width=\linewidth]{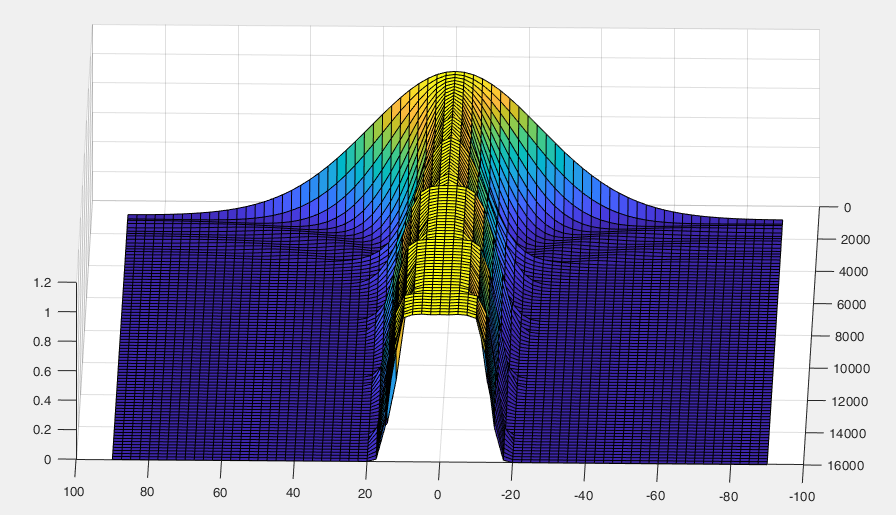}
	\caption{Beam-shape at elevator from $0^o$ to $90^o$, azimuth=$0^o$, $180^o$, frequency from 1Khz to 16Khz. The Aperture is almost constant over frequencies.}
	\label{fig:fig9}
	\end{figure}
\end{itemize} 

\begin{exa}
design the beam pattern with open-angle having 2 levels of gain:	
\end{exa} 
\begin{itemize}
\item Step 1: Define the expected beam-pattern (Figure.\ref{fig:fig10})\\ 
$\theta \geq \theta_{C1}$: gain 1  \\
$\theta_{C1} \geq \theta \geq \theta_{C2}$: gain $1/10$ (reduce 20 dB) \\
Others: gain 0
	\begin{figure}[h!]
	\includegraphics[width=\linewidth]{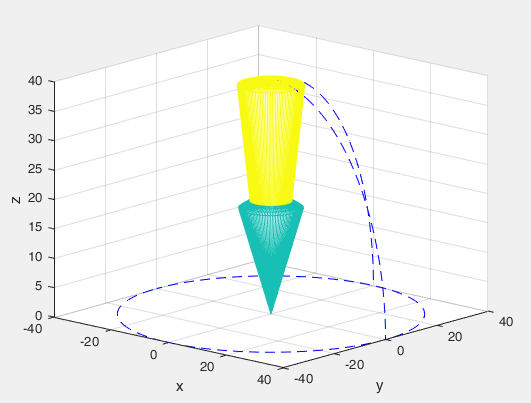}
	\caption{Expected Beam-Pattern in dB scale, maximum dB scale is 40 dB, equivalence with gain 1.}
	\label{fig:fig10}
	\end{figure}
\item Step 2: define a surface function in spherical coordinate.  
\[
b_R( \theta , \phi) \ = {
	\cases{1, \; \hbox{ $\theta \leq \theta_{C1}$,} \  \cr
		1/10, \;  \hbox{$\theta_{C1} < \theta \leq \theta_{C2}$,}  \ \cr
		0, \;     \hbox{Otherwise.}
}}
\]
$\theta_{C1}$  : Threshold 1, define the beam-width 40dB.\\
$\theta_{C2}$  : Threshold 2, define the beam-width 20dB.
\item 	Step 3: Convert $b_R(\theta,\phi)$  to Cartesian coordinate ($Rsin\theta= \sqrt{u^2+v^2}, R = {Nd_Hf \over c} $):
\[
b_f(u,v) \ = {
	\cases{1, \; \hbox{$\sqrt{u^2+v^2} \leq R\sin\theta_{C1}$} \  \cr
		1/10, \;    
		\hbox{$R\sin\theta_{C1} < \sqrt{u^2+v^2} \leq R\sin\theta_{C2}$}  \ \cr
		0, \;     \hbox{Otherwise.}
}}
\]\\
The shape in Cartesian Coordinate ( with $N$=100, $d_H$=0.015m): Figure.\ref{fig:fig11}.
	\begin{figure}[h!]
	\centering	
	\includegraphics[width=\linewidth]{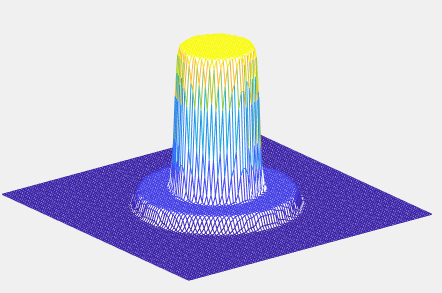}
	\caption{Expected Beam-Pattern transforms to Cartesian Coordinate at 16Khz.}
	\label{fig:fig11}
	\end{figure}

\item Step 4: The weight is achieved by IDFT, Figure.\ref{fig:fig12}.\\ 
	\begin{figure}[h]
	\includegraphics[width=\linewidth]{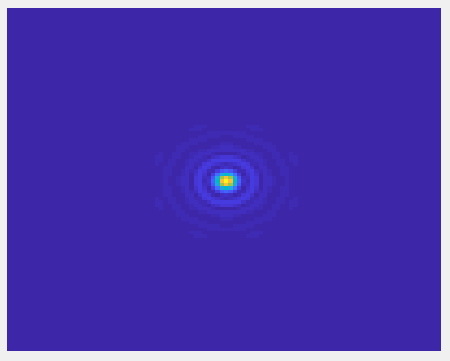}
	\caption{Weight at 16Khz.}
	\label{fig:fig12}
	\end{figure}\\
Real beam pattern at 16Khz are depicted in Figure.\ref{fig:fig13}.
The cross-cut of beam patterns versus frequencies is depicted in Figure.\ref{fig:fig14}.\\

	\begin{figure}[h!]
	\includegraphics[width=\linewidth]{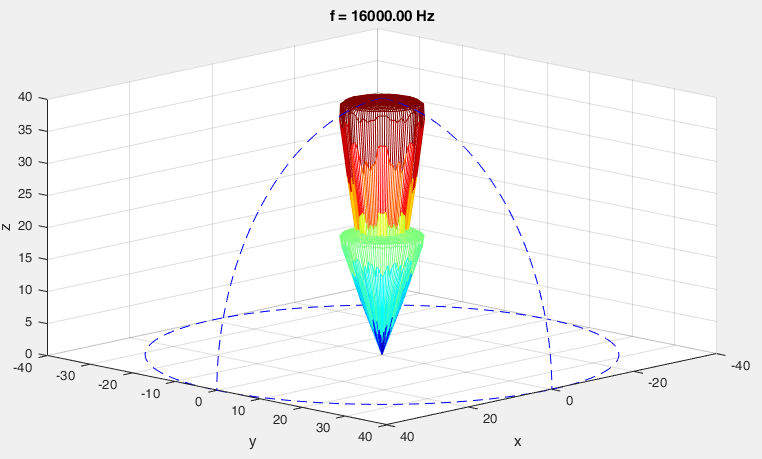}
	\caption{Beam-Pattern at 16Khz.}
	\label{fig:fig13}
	\end{figure}

	\begin{figure}[h!]
	\includegraphics[width=\linewidth]{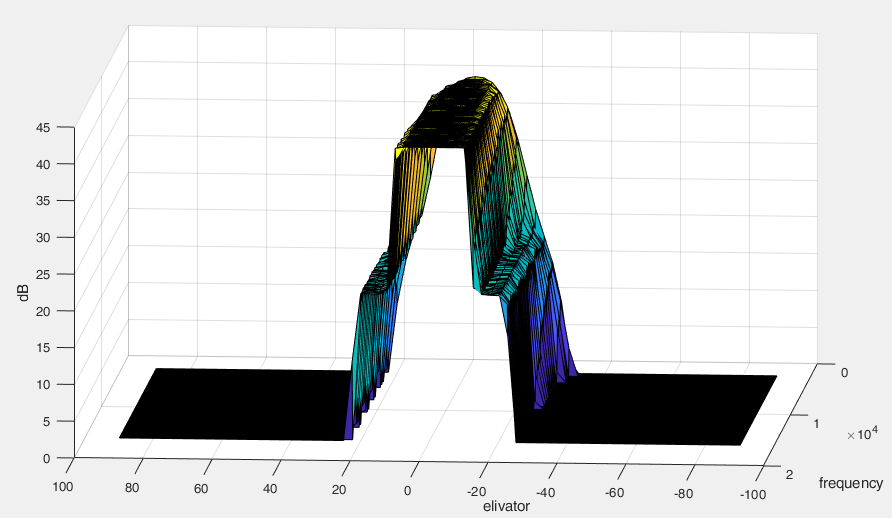}
	\caption{Beam-shape at elevator from $0^o$ to $90^o$, azimuth=$0^o$, $180^o$, frequency from 2.5Khz to 16Khz.}
	\label{fig:fig14}
	\end{figure}
\end{itemize} 	
\begin{exa}
 Design beam pattern with 2D-Sinc shape:
\end{exa}

\begin{itemize}
\item Step 1: Define the expected beam-pattern, Figure.\ref{fig:fig15}.\\ 
\[
B( \theta , \phi) \ = {
	\cases{\vert {{\sin(\alpha\pi\theta)} \over {\alpha\pi\theta} }\vert, \; \hbox{    $\theta > 0$, $\alpha$ is a constant,} \  \cr
		1, \; \hbox{ $\theta = 0$.}
}}
\]
	\begin{figure}[h!]
	\centering
	\begin{subfigure}[b]{0.4\linewidth}	
	\includegraphics[width=\linewidth]{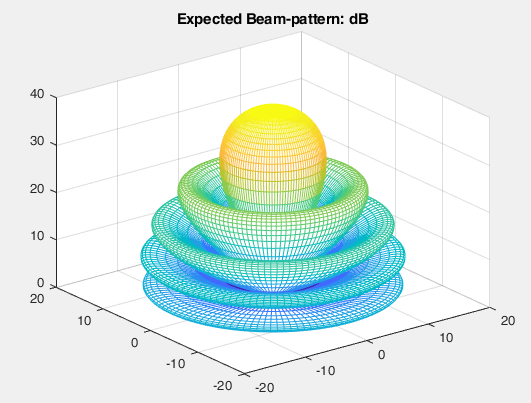}
	\caption{}
    \end{subfigure}
	\begin{subfigure}[b]{0.4\linewidth}
		\includegraphics[width=\linewidth]{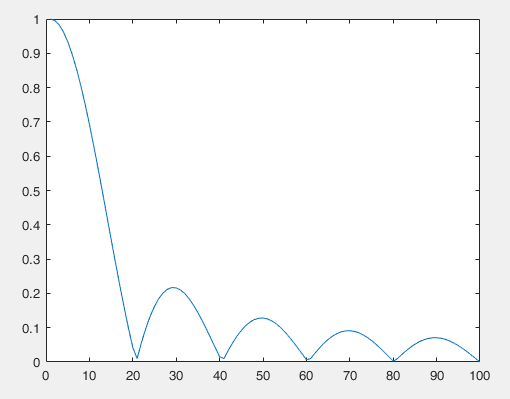}
		\caption{ cross-cut.}
		
	\end{subfigure}
    \caption{(a): Expected beam pattern (in dB scale). (b): Cross-cut of beam pattern.}
    \label{fig:fig15}
    \end{figure}

\item Step 2: define a surface function in spherical coordinate.  \\
\[
b_R( \theta , \phi) \ = {
	\cases{\vert {\sin(\alpha\pi\theta) \over \alpha\pi\theta }\vert, \; \hbox{    $\theta > 0$, $\alpha$ is a constant,} \  \cr
		1, \;  \hbox{$\theta = 0$}
}}
\]
\\

\item Step 3: Convert $b_R(\theta,\phi)$  to Cartesian coordinate ( With $N$=200, $d_H$=0.01m), Figure.\ref{fig:fig17}. \\
	\begin{figure}[h!]
	\includegraphics[width=\linewidth]{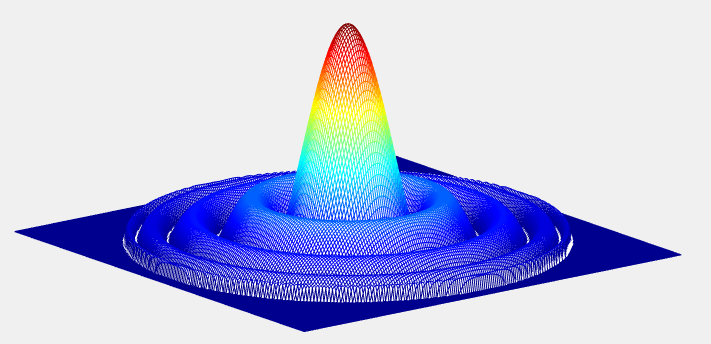}
	\caption{Cartesian Coordinate:Beam Pattern at 16Khz.}
	\label{fig:fig17}
\end{figure}\\
The relationship between Sphere coordinate and Cartesian coordinate: $\theta = \arcsin(\frac{\sqrt{u^2+v^2}}{R})$.\\
Replace $R$ from (\ref{Radius:Dis}) we get:
$\theta = \arcsin(\frac{c\sqrt{u^2+v^2}}{ fNd_H})$\\
Then,

\[
b_f( u, v) \ = {
	\cases{\left\vert 
		{\sin(\alpha\pi\arcsin({c\sqrt{u^2+v^2} \over fNd_H})) 
		\over 
	     \alpha\pi\arcsin({c\sqrt{u^2+v^2} \over fNd_H}) }
		\right\vert, \; \ \cr
		\qquad \hbox{$u,v \ne 0$, $\sqrt{u^2+v^2} \leq {fNd_H \over c}$,} \  \cr
		1, \;  \hbox{ $u=v= 0$,}  \ \cr
		0, \;  \hbox{  Otherwise.}
}}
\]
\\

\item Step 4: 
The weight is achieved by IDFT, Figure.\ref{fig:fig18}. 
Beam pattern at 16Khz is showed in Figures \ref{fig:fig19}.
\begin{figure}[h!]
	\includegraphics[width=\linewidth]{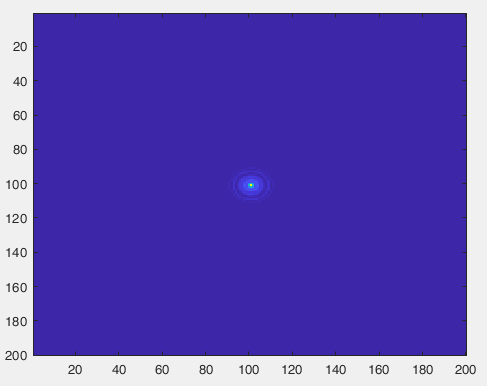}
	\caption{Magnitude of Weights at 16Khz.}
	\label{fig:fig18}
\end{figure}
	\begin{figure}[h!]
	\includegraphics[width=\linewidth]{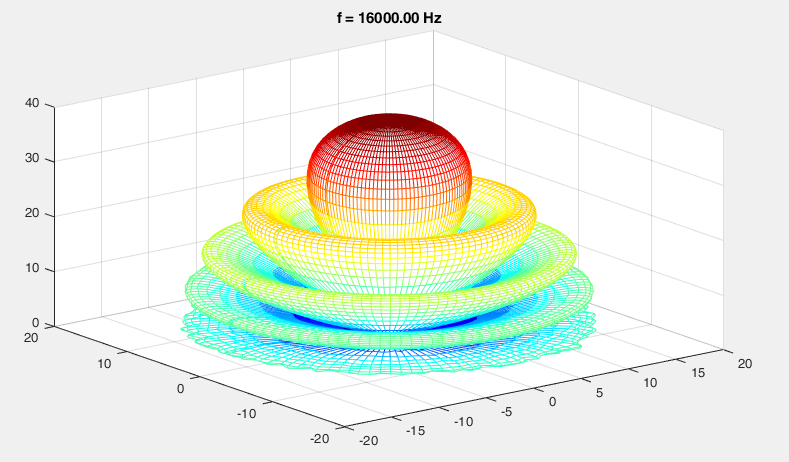}
	\caption{Beam-Pattern at 16Khz (in dB scale).}
	\label{fig:fig19}
	\end{figure}\\
Cross-cut of beam pattern from 2.5Khz to 16kHz, Figure.\ref{fig:fig20}.
	\begin{figure}[h!]
	\includegraphics[width=\linewidth]{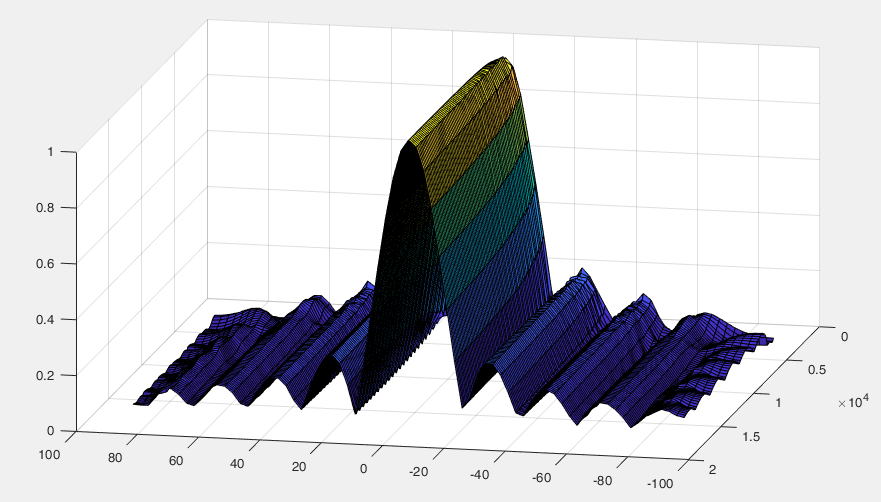}
	\caption{Beam-shape at elevator from $0^o$ to $90^o$, azimuth=$0^o$, $180^o$, frequency from 2.5Khz to 16Khz.}
	\label{fig:fig20}
	\end{figure}
\end{itemize} 
	
\section{Conclusions}
Using the relationship of beam pattern and Fourier transform, beam pattern’s geometry translation and transformation, we propose a method to design the arbitrary beam pattern for planar array and line array. We also explain the difference between the expected beam pattern and the real beam pattern. It results from the effect of geometry’s deformation during transforming the beam pattern to the Cartesian coordinate. The computation time of new method is much less than optimization methods because it only uses coordinate transformation and Invert Fourier transform for computation.
 
So far, we all know the spatial sampling theory states: inter-distance of sensors should be less than half of wavelength
$${d_H < {\lambda \over 2}}.$$ 
However, from (\ref{Dis:Condition}) we could observe: to avoid deformation of beam pattern 
$${d_H \leq {{(N-1)\lambda} \over {2N}}}.$$
This an upper bound for the inter-distance of sensors is tighter than the spatial sampling theory's upper bound which is true for continuous sensor array. The new upper bound is close to spatial sampling theory's upper bound if $N$ is big enough. In general, we could use the new upper bound for discrete sensor array.

If we are interested in a lower bound for inter-distance of sensors, from (\ref{Dis:Condition}) we have,
$${d_H \geq {\lambda \over N}},$$  
the inter-distance of sensors should be greater than or equal the wavelength devides by the number of sensors in one dimension (vertical or horizontal) of planar array.

\end{document}